\documentclass[11pt]{article}
\usepackage{amsmath, amssymb, amsthm}
\usepackage{graphicx}
\usepackage{algorithmic}
\usepackage{algorithm}
\usepackage{graphicx}
\usepackage{cite}

\begin{document}
	
	\title{HistoSPACE: Histology-Inspired Spatial Transcriptome Prediction And Characterization Engine}
\author{Shivam Kumar$^{\dagger}$ and Samrat Chatterjee$^{\dagger}$\thanks{Corresponding author: Samrat Chatterjee, e-mail: samrat.chatterjee@thsti.res.in}\\
\\
$^{\dagger}$ Complex Analysis Group, \\Translational Health Science and Technology Institute, \\ NCR Biotech Science Cluster, \\ Faridabad-Gurgaon Expressway, Faridabad-121001, India\\
India.}

\date{}
\maketitle

\begin{abstract}

Spatial transcriptomics (ST) enables the visualization of gene expression within the context of tissue morphology. This emerging discipline has the potential to serve as a foundation for developing tools to design precision medicines. However, due to the higher costs and expertise required for such experiments, its translation into a regular clinical practice might be challenging. Despite the implementation of modern deep learning to enhance information obtained from histological images using AI, efforts have been constrained by limitations in the diversity of information. In this paper, we developed a model, HistoSPACE that explore the diversity of histological images available with ST data to extract molecular insights from tissue image. Our proposed study built an image encoder derived from universal image autoencoder. This image encoder was connected to convolution blocks to built the final model. It was further fine tuned with the help of ST-Data. This model is notably lightweight in compared to traditional histological models. Our developed model demonstrates significant efficiency compared to contemporary algorithms, revealing a correlation of 0.56 in leave-one-out cross-validation. Finally, its robustness was validated through an independent dataset, showing a well matched preditction with predefined disease pathology.\\

\end{abstract}

\section{Introduction} \label{sec:introduction}

In modern biomedicine, understanding the relationship between spatial organization and gene expression within the tissue is of great significance \cite{parada2004tissue,behanova2022spatial}. Traditional transcriptome analysis using sequencing has greatly expanded our knowledge of genetic profiles. However, they often fail to capture the spatial heterogeneity that characterizes complex biological systems \cite{liao2022novo,saul2023spatial}. This limitation is obvious when investigating diseases with heterogeneity. For example, in cancer, the tumor microenvironment (TME) and cell-to-cell interactions play pivotal roles in disease progression and therapeutic responses \cite{hsieh2022spatial,quail2013microenvironmental,de2023evolving}. To understand such phenomena, spatial transcriptomics (ST) has emerged as a groundbreaking methodology for bridging this fundamental gap. It captures spatially resolved transcript expression using bar-coded DNA, which distinguishes different spots in the tissue, similar to the Cartesian plane. Each spot position contains two to dozens of cells varying with different ST technologies \cite{song2021dstg,yue2023guidebook}. So, we have a gene expression profile vector for hundreds of spots, collectively representing multiple points across the entire tissue. The data collected from spatial transcriptomics allows us to investigate gene expression patterns within the structural context of tissues \cite{rao2021exploring,walker2022deciphering,d2022spatially}. This approach has improved our understanding of various biological phenomena, from the developmental processes of spermatids \cite{chen2021dissecting} to the pathological mechanisms underlying cancer and other metabolic disease like diabetes \cite{williams2022introduction,ya2023application}.

Notwithstanding the great progress in spatial transcriptomics, its complete potential remains underexplored. The constraints primarily arise from two critical factors: The prohibitive cost associated with experimentation and the substantial expertise required \cite{liu2022analysis}. In contrast, whole-slide images (WSIs) offer a more cost-effective and readily accessible alternative. Histopathology analysis of tumor biopsy sections remains a cornerstone of clinical oncology. Such as visual inspection of hematoxylin and eosin (H\&E)-stained WSIs provide valuable insights into tissue morphology. Histopathology is a routine practice in clinical settings for accurate diagnoses and grading \cite{ellison2011histopathological,baxi2022digital}. Notably, WSIs have demonstrated a high correlation with bulk gene expression, serving as the foundation model for predicting gene expression profiles \cite{mondol2023hist2rna,zeng2022spatial,comiter2023inference,wang2021predicting}.

In the contemporary biomedical research landscape, combining deep learning (DL) techniques with histology has emerged as a formidable tool. The application ranges from deciphering intricate cellular structures, detecting subtle pathological changes, and identifying biomarkers indicative of disease states, all with high accuracy \cite{badea2020identifying,van2021deep,wu2021precision,barisoni2020digital}. Such advancements have augmented the capabilities of pathologists by automating tasks such as mitosis detection, quantifying tumor immune infiltration, classifying cancer subtypes, and grading tumors \cite{wang2023generalizable,ankitha2023brain,wetstein2022deep,chen2020deep,abousamra2022deep}. The recent successes of DL in biomedicine provided us with a solid foundation to bridge the gap between molecular signatures and WSIs. The integration of molecular profiling is important in deciphering the intricate tissue heterogeneity within the context of disease \cite{ya2023application}. This amalgamation of cutting-edge technologies has the potential to revolutionize drug discovery by harnessing histology for the identification and development of personalized medicine \cite{zhang2022clinical}. Deep learning algorithms can be trained to process spatial transcriptomic data, allowing for the extraction of biologically meaningful information from the complex datasets generated by this technology \cite{zeng2022statistical}. By integrating these two powerful approaches, researchers can discern patient-specific variations in tissue composition and gene expression. The ability to glean detailed molecular insights from individual patient tissue samples, coupled with the predictive capabilities of deep learning, promises to accelerate the development of personalized therapeutics. By tailoring drug candidates to the unique molecular signatures of each patient, we stand on the precipice of a paradigm shift in healthcare, where treatments are more effective and safer, with reduced adverse effects.

Some initial work was done to integrate ST with histology. Among the available approaches, some rely on deploying complex neural network architectures. While promising, these models introduce certain technical challenges. Their computational demands often necessitate significant hardware resources, potentially limiting their practical utility, especially when working with large-scale datasets or in resource-constrained environments. Additionally, the intricate nature of these neural architectures can lead to challenges in generalization across different datasets and biological contexts. This work aims to develop a novel autoencoder-based algorithm, HistoSPACE, for gene expression prediction from histological images. This algorithm seeks to balance model complexity and accuracy, offering a compact yet powerful solution for spatial transcriptomic profiling. 

The main contribution of this work is as follows,
\begin{enumerate}
	\item Developed an auto-encoder base image feature extractor,
	\item Proposed a methodology for gene expression prediction from histology images.
\end{enumerate}

\section{Literature review}

We reviewed the literature on auto-encoder models for images and spatial transcriptomics pattern prediction from H\&E images. These elements serve as the foundation of our study, guiding our exploration of existing research.

\subsection{Autoencoders for Images}

Autoencoders have emerged as powerful tools for image analysis and feature extraction in various domains, including the biological sciences. They have demonstrated their versatility in capturing intricate patterns within complex biological images, enabling researchers to gain deeper insights into various biological phenomena \cite{silva2020pre,ponzio2019dealing}. Autoencoders have proven valuable in biological imaging for image denoising and restoration tasks. By training autoencoders on noisy or degraded biological images, researchers can effectively remove artifacts and enhance the image quality, thus improving the accuracy of subsequent analyses \cite{thomas2022bio,gondara2016medical}.

Furthermore, autoencoders have found applications in feature learning and representation. In cellular and tissue imaging, these models can identify relevant features and structures, such as organelles, nuclei, or cell types \cite{singha2023alexsegnet,xu2015stacked,sun2022improving}.

Another noteworthy application is in image classification as a transfer learning approach, where autoencoders are employed as feature extractors of images. Connecting the encoder part of these models is a powerful solution in biomedical image classification, such as predicting Parkinson \cite{hema2022prediction}, COVID-19 diagnosis \cite{addo2022evae}, lung nodule classification \cite{mao2018feature}. These work as foundational models to mitigate the challenge of limited labelled data and enhance the performance of classification models.

\subsection{Expression prediction}

Several established approaches have exhibited promising outcomes in the domain of predicting gene expression from histology images, encompassing methods such as HE2RNA \cite{schmauch2020deep}, ST-Net \cite{he2020integrating}, HisToGene \cite{pang2021leveraging}, hist2rna \cite{mondol2023hist2rna}, Hist2ST \cite{zeng2022spatial}, and others\cite{comiter2023inference,wang2021predicting}. Among them, ST-Net and His2ST have emerged as two major techniques for forecasting spatially resolved expression patterns from H\&E images. These methods treat the task of expression prediction as regression problems, employing feed-forward training mechanisms. ST-Net incorporates a densenet121 image encoder followed by a fully connected layer, while HisToGene harnesses a vision transformer backbone with an expanded field of view.

While methods targeting tissue-level bulk expression prediction generally achieve good correlation, they are typically unable to generate spatially resolved expression profiles (as observed with HE2RNA). On the other hand, existing methods capable of generating spatially resolved expression predictions have limitations, including a lack of external evaluation, restrictions on the predicted gene sets, and susceptibility to overfitting \cite{wang2021predicting}.

\section{Methodology}

\begin{figure}[H]
	\centerline{\includegraphics[width=\textwidth]{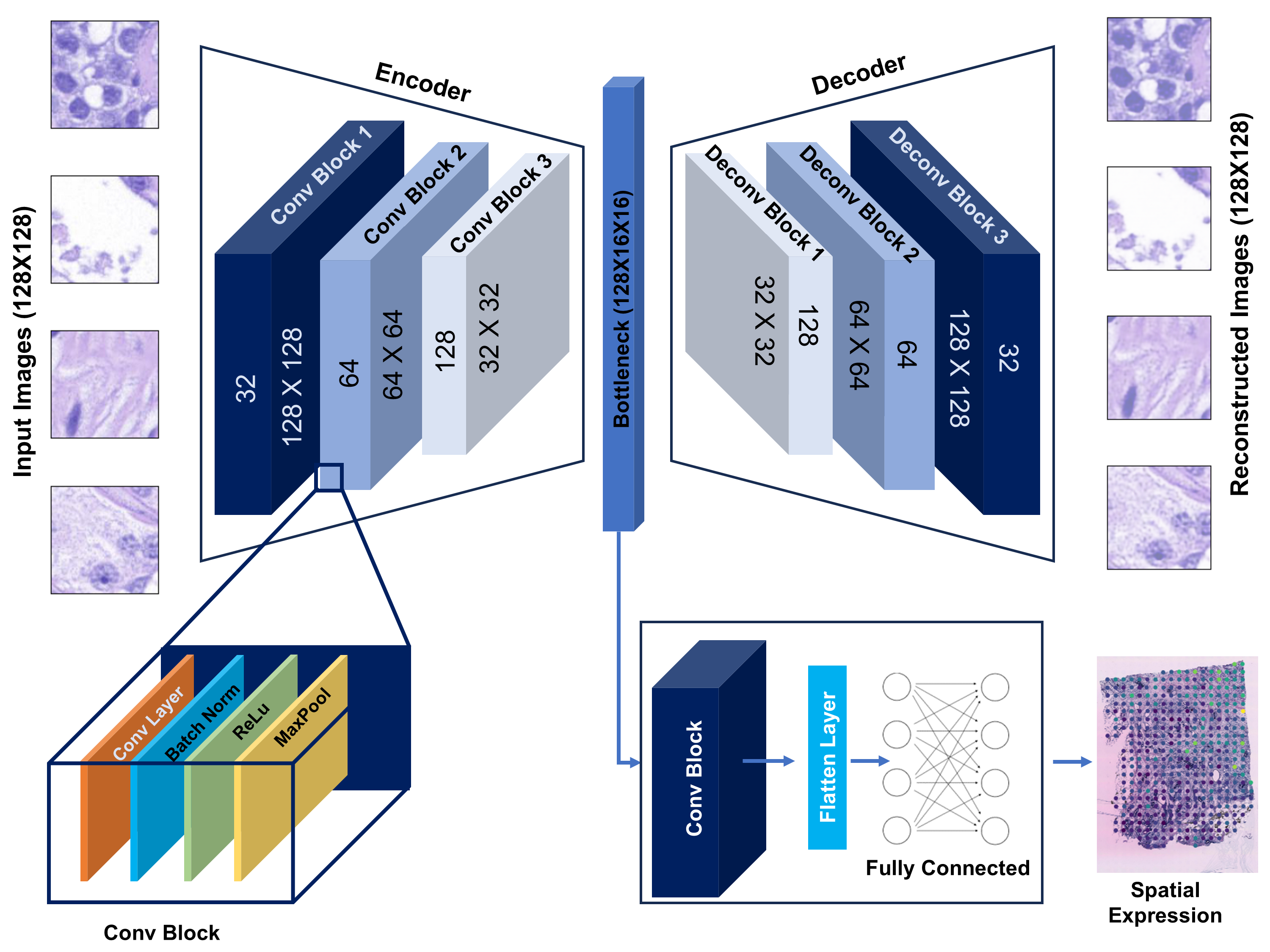}}
	\caption{{\bf Brain of HistoSPACE:}  This is a complete framework consisting of two major components, an image autoencoder and an expression prediction model. The upper portion shows the image encoder-decoder network. The lower left shows the components of the convolution block, which are convolution, batch normalization, Relu and maximum pooling layers. The lower right shows, where encoder is connected with convolution blocks to predict the spatial expression.}
	\label{fig1}
\end{figure}

The proposed algorithm presents a multi-model framework designed to enhance the interpretability and performance of spatial transcriptomics prediction (See Figure \ref{fig1}). Initially, an end-to-end image autoencoder is trained on distinct and independent breast cancer datasets, allowing it to capture intricate data representations. 

The autoencoder's encoder component is leveraged as a feature extractor in this algorithm's final stage. These extracted features were subsequently utilized by adding custom convolution operations followed by fully connected layers for gene expression prediction. This multi-step approach enhances the predictive accuracy and sheds light on the critical features and patterns associated with the expression.

\subsection{Image Autoencoder}

Building an autoencoder involves designing an architecture with an encoder to compress input data and a decoder to reconstruct it \cite{pinaya2020autoencoders}. The encoder consists of three convolution blocks responsible for extracting hierarchical features. The shape of the first convolution block is similar to the size of an input image. The output channels are 32 - 64 -128 in respective blocks. This increased depth helps to capture complex and abstract features as we move towards deep layers \cite{vincent2010stacked}. The convolution block has a convolution layer, batch normalization, ReLU activation, and max-pooling layer. Batch normalization helps to stabilize and accelerate the model training, and the max-pooling layer is used to down-sample the spatial dimensions \cite{luo2017convolutional,li2018discriminatively}. The decoder architecture also follows the same depth pattern but in reverse order. An up-sampling layer replaces its convolution layer, and the last layer is sigmoidal, which helps to stabilize the pixel value during image reconstruction \cite{cheng2018deep}. During training, the model minimizes the reconstruction loss, which measures the difference between the input and the output, a mean squared error (MSE) in our case. The training aims to optimize the model’s parameters with minimal loss, ensuring that the autoencoder can faithfully reconstruct the input data.
\begin{equation}
	\text{Autoencoder Loss } (L_{AE}) = \sum_{i=1}^{N} ||x_i - \hat{x}_i||^2,
\end{equation}
where $x_i$  represents the input image, and $\hat x_i$ represents the image reconstructed by the autoencoder. The number of samples is represented by $N$.

\subsection{Building expression prediction model}

The final model relies on the encoder component of our trained autoencoders. This component is a feature extractor of histology images for the final model. We added new layers to the encoder and proposed a gene expression prediction model. The extracted features were mapped with corresponding gene expression data and only trained the additional layer in the model. The final trained model is HistoSPACE, which stands for Histology-Inspired Spatial Transcriptome Prediction And Characterization Engine. The HistoSPACE architecture is designed with careful consideration of both the encoder’s output and the requirements of the gene expression prediction task. The final model was tested by adding a few convolution blocks. The Convolution block has a convolution, batch normalization, ReLU activation, and max-pooling layer. These components work together to ensure that relevant spatial information is captured. These convoluted features, denoted as $Z$, were then connected to fully connected layers to map the extracted features to gene expression levels. The nodes in the last fully connected layers equal the number of predicted genes, catering to the image regression problem. 
\begin{equation}
	\hat Y = F_{fc}(Z),
\end{equation}
here, $Z$ represents the encoded features, and $F_{fc}$ or fully connected represents the custom layers responsible for making the gene expression predictions. The purpose of these $F_{fc}$ layers is to translate these features into predictions of gene expression, denoted as $\hat Y$.

The model was finalized by systematically increasing the complexity of the added layers. Specifically, we augment the model by introducing an additional convolutional block and increasing the dimensionality of the fully connected layers. These architectural modifications enhance the model's capacity to learn intricate patterns in the gene expression data.

\begin{algorithm}
	\caption{Predicting Spatial Gene Expression from Histology Images}
	\begin{algorithmic}
		\STATE \textbf{Input:} Histology images
		\STATE \textbf{Output:} Predicted gene expression\\
		\STATE \textbf{Step 1:} Preprocessed Histology Images
		\begin{itemize}
			\item Convert the histology images into small patches.
		\end{itemize}
		
		\STATE \textbf{Step 2:} Train an Image Autoencoder
		\begin{itemize}
			\item Initialize the image autoencoder.
			\item Train the autoencoder using the small image patches.
			\item Perform multiple calibration iterations to fine-tune the autoencoder.
			\item Obtain the final autoencoder model.
		\end{itemize}
		
		\STATE \textbf{Step 3:} Extract Encoder from Autoencoder
		\begin{itemize}
			\item Extract the encoder part from the final autoencoder.
		\end{itemize}
		
		\STATE \textbf{Step 4:} Build the Gene Expression Predictor
		\begin{itemize}
			\item Add a convolutional layer after the encoder.
			\item Add two fully connected (FC) layers to the model.
			\item Configure the model for gene expression prediction.
		\end{itemize}
		
		\STATE \textbf{Step 5:} Training
		\begin{itemize}
			\item Train the model using the histology images and corresponding gene expression data.
		\end{itemize}
		
		\STATE \textbf{Step 6:} Gene Expression Prediction
		\begin{itemize}
			\item Use the trained model to predict gene expression from new histology images.
		\end{itemize}		
	\end{algorithmic}
\end{algorithm}

\subsection{Selecting the loss functions}

The objective of this model is to predict gene expression from histology images. Predicting continuous value from a model comes under regression tasks, where selecting an appropriate loss function is pivotal to the model’s performance. The usual choices of loss function for such task are Mean Squared Error (MSE) and Root Mean Squared Error (RMSE), calculated as the average of the squared differences between predicted and actual values, which strongly emphasises outliers. While RMSE, the square root of MSE, measures the average error with the same emphasis on outliers, it is more interpretable as it is in the same units as the target variable. However, MSE and RMSE can be sensitive to outliers, which may be present in real-world gene expression data.

In contrast, the Huber loss function offers a robust alternative. It combines MSE and Mean Absolute Error (MAE) characteristics. Employing a squared loss for more minor errors and an absolute loss for more significant errors, where a hyperparameter controls the balance, often denoted as $\delta$. This inherent adaptability allows the Huber loss to effectively handle outliers, making it less prone to their influence than MSE or RMSE.
\begin{equation}
	{MSE}(x_i,\hat{x}_i) = \sum_{i=1}^{N} ||x_i - \hat{x}_i||^2,
\end{equation}
where $x_{i}$ is input and $\hat x_{i}$ is the predicted expression, respectively. $N$ is the number of samples.
\begin{equation}
	{RMSE}(x_i,\hat{x}_i) = \sqrt{\sum_{i=1}^{N} ||x_i - \hat{x}_i||^2	},
\end{equation}
where $x_{i}$ is input and $\hat x_{i}$ is the predicted expression, respectively. $N$ is the number of samples.
\begin{equation}
	H_\delta(y, f(x)) =
	\begin{cases}
		\frac{1}{2}(y - f(x))^2, & \text{if } |y - f(x)| \leq \delta, \\
		\delta (|y - f(x)| - \frac{1}{2}\delta), & \text{otherwise,}
	\end{cases}
\end{equation}
where $y$ is input and $f(x)$ is the predicted expression respectively. The smoothening parameter is defined as  $\delta$.

\subsection{Performance evaluation using correlation}

To evaluate the performance of models, we are calculating pearson correlation coefficient ($r$) between predicted expression profile with actual/true expression profile of a sample,

\begin{equation}
	{r} = \frac{{}\sum_{i=1}^{n} (x_i - \overline{x})(y_i - \overline{y})}
	{\sqrt{\sum_{i=1}^{n} (x_i - \overline{x})^2  \sum_{i=1}^{n}(y_i - \overline{y})^2}}
\end{equation}

where $x$ and $y$ represent true and predicted expression values respectively, $i$ has maximum value of $n$ which is total number of genes for that sample. 

Final performance is reported by taking average of $r$ for all such samples in consideration.

\section{Experiments and Evaluations}

\subsection{Datasets}

This study uses three publicly available datasets, each described below.

\begin{figure}[H]
	\centerline{\includegraphics[width=\columnwidth]{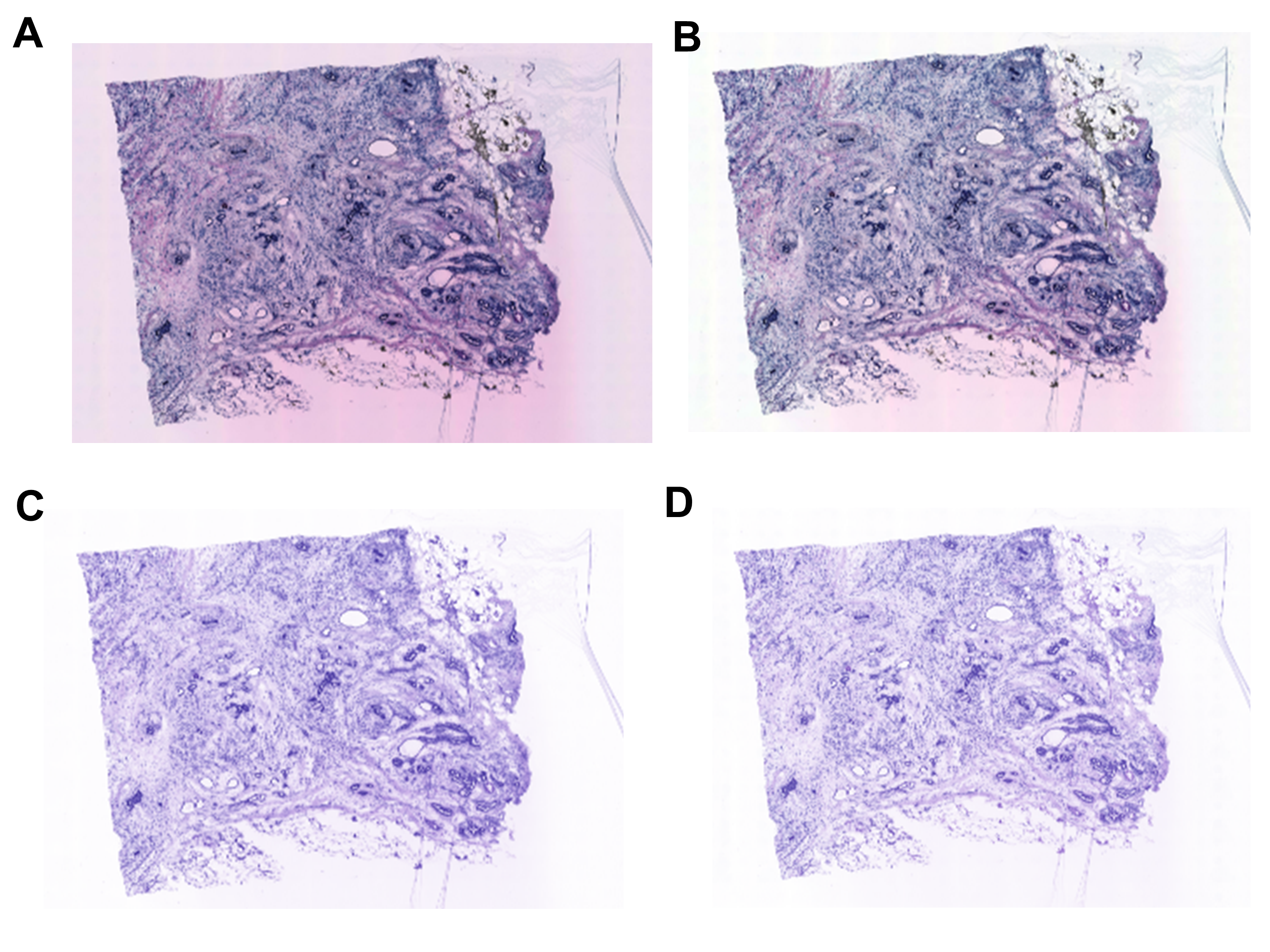}}
	\caption{{\bf Effect of image normalization.}
	(A) Raw image. (B) Color cast removed. (C) Stain normalized with the reference image. (D) Combine the effect of color cast removal and stain normalization.}
	\label{fig2}
\end{figure}

\subsubsection{ICIAR Dataset } 

This dataset contains microscopy images labelled as normal, benign, in situ carcinoma, or invasive carcinoma according to the predominant cancer type in each image \cite{araujo2017classification}. Two medical experts performed the annotation, and the study did not consider images with disagreement. Each category has 100 images for training and 25 for testing, leading to total of 400 and 100, respectively. 

\begin{figure}[H]
	\centerline{\includegraphics[width=\textwidth]{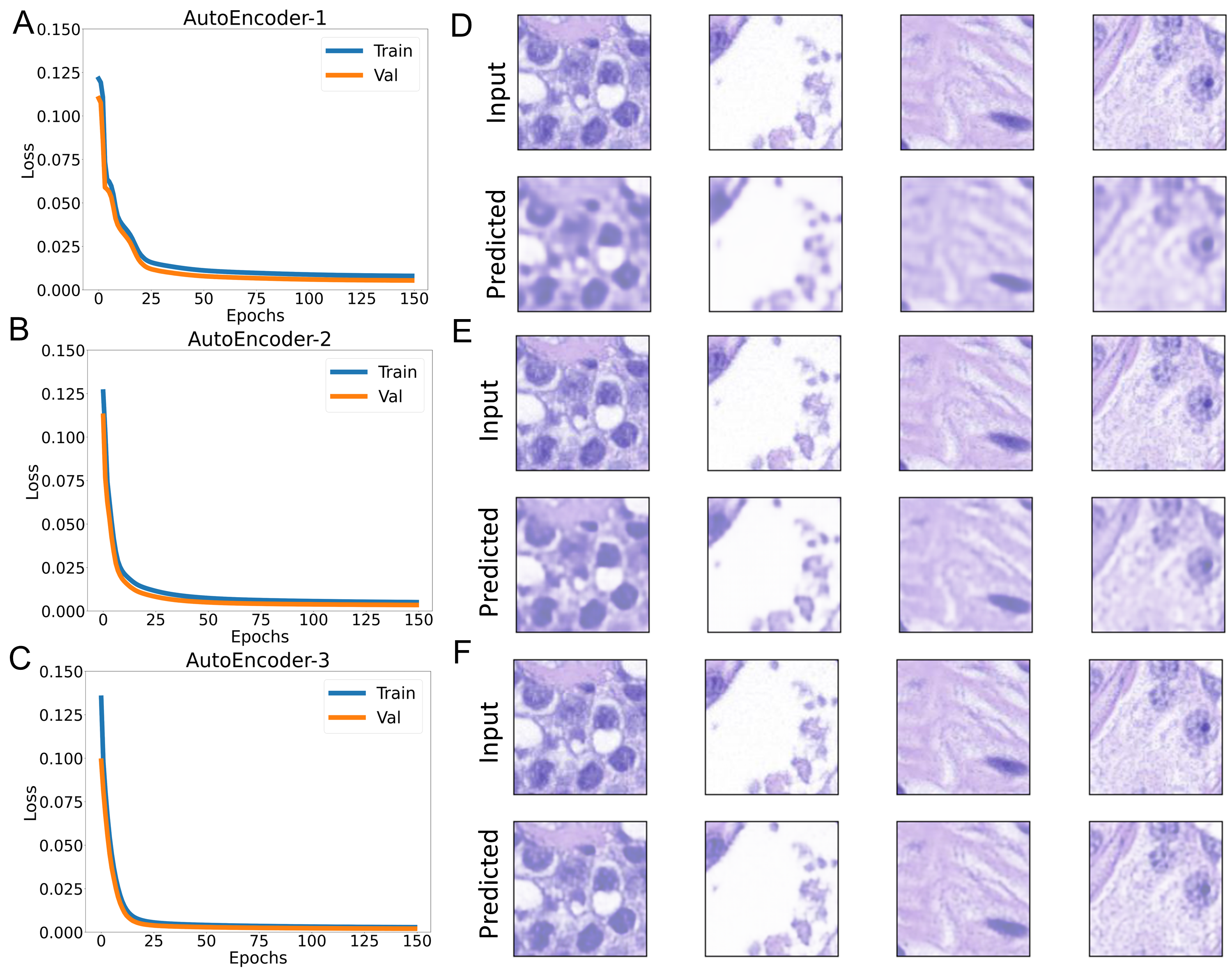}}
	\caption{{\bf Building the autoencoder model.}
		(A-C) The learning curve of training and test data for the three models, basic to advanced. (D-F) The prediction images from the testing tile set for the basic to advanced models show better imaging features captured.}
	\label{fig3}
\end{figure}

\subsubsection{STNet Dataset } 

This study encompasses 23 patients diagnosed with breast cancer \cite{he2020integrating}. This dataset has three microscope images of H\&E-stained tissue slides for each patient, coupled with corresponding spatial transcriptomics data. Within a single tissue section, spatial transcriptomics quantifies RNA expression within spots featuring 100 $\mu m$ diameter, arranged in a grid with a center-to-center distance of 200 $\mu m$.

\subsubsection{HER2 Dataset } 

To capture the model's generalization and present a case study for precision oncology, we have used this dataset \cite{andersson2021spatial}. Spatial transcriptome profiles were generated from eight breast cancer patients using the $10\times$ Visium platform.

\subsection{Image standardization technique}

Stain normalization plays a pivotal role in histology for deep learning due to its importance in enhancing the quality and consistency of histological images. The primary reason for stain normalization is to mitigate the variability in staining procedures and image acquisition techniques commonly encountered in histology. These variations can introduce unwanted artifacts, such as color casts, shadows, and gradients, hindering deep learning models' accurate interpretation of tissue features. The raw image without processing often exhibits shadows and unnecessary color gradients, which can be misleading for deep learning algorithms (see Figure \ref{fig2}A). Removing color cast helps eliminate additional background color, improving the image's clarity and reducing noise (see Figure \ref{fig2}B). Stain normalization further enhances image quality by freeing it from staining artifacts, ensuring consistent color representation across different samples (see Figure \ref{fig2}C). Combining color cast removal and stain normalization produces well-prepared images for deep learning algorithms  (see Figure \ref{fig2}D). These processed images provide a clean and uniform representation of tissue structures, facilitating the algorithms' accurate feature extraction and analysis.

\subsection{Image feature extractor from autoencoder}

We present the results of three different iterations of autoencoder models ({\bf see supplementary figure s1}), denoted as AutoEncoder 1, AutoEncoder 2, and AutoEncoder 3. Each variant incorporates unique architectural elements, as illustrated by Figure \ref{fig3}. Our evaluation of these variations is rooted in analysing their respective learning curves and the overall quality of the images they generate.
The input to the autoencoder comprises non-overlapping image tiles, each measuring 128X128 pixels. These tiles are extracted from H\&E (Hematoxylin and Eosin) images from the ICIAR Dataset. Using non-overlapping image tiles allows us to comprehensively explore the autoencoders' capabilities in reconstructing the image features.

AutoEncoder-1 represents the baseline model with only max-pooling layers. While this model achieved an error of 0.0048 and exhibited some pattern recognition, the generated images appeared hazy and needed more fine details.

In AutoEncoder-2, we introduced an additional ReLU activation layer to enhance the model's ability to capture nonlinear patterns. This modification decreased the error from 0.0048 to 0.0034 and improved pattern recognition, but the generated images retained some blur effect.

AutoEncoder-3 represents the most advanced version of our autoencoders, featuring both ReLU activation layers and batch normalization. Notably, the learning curve for this model stabilized over 150 epochs, showcasing a good learning behavior characterized by progressively decreasing error rates led to 0.002. Moreover, the images generated by AutoEncoder-3 demonstrated better clarity and finer details than the previous models.

\begin{figure}[H]
	\centerline{\includegraphics[width=0.6\columnwidth]{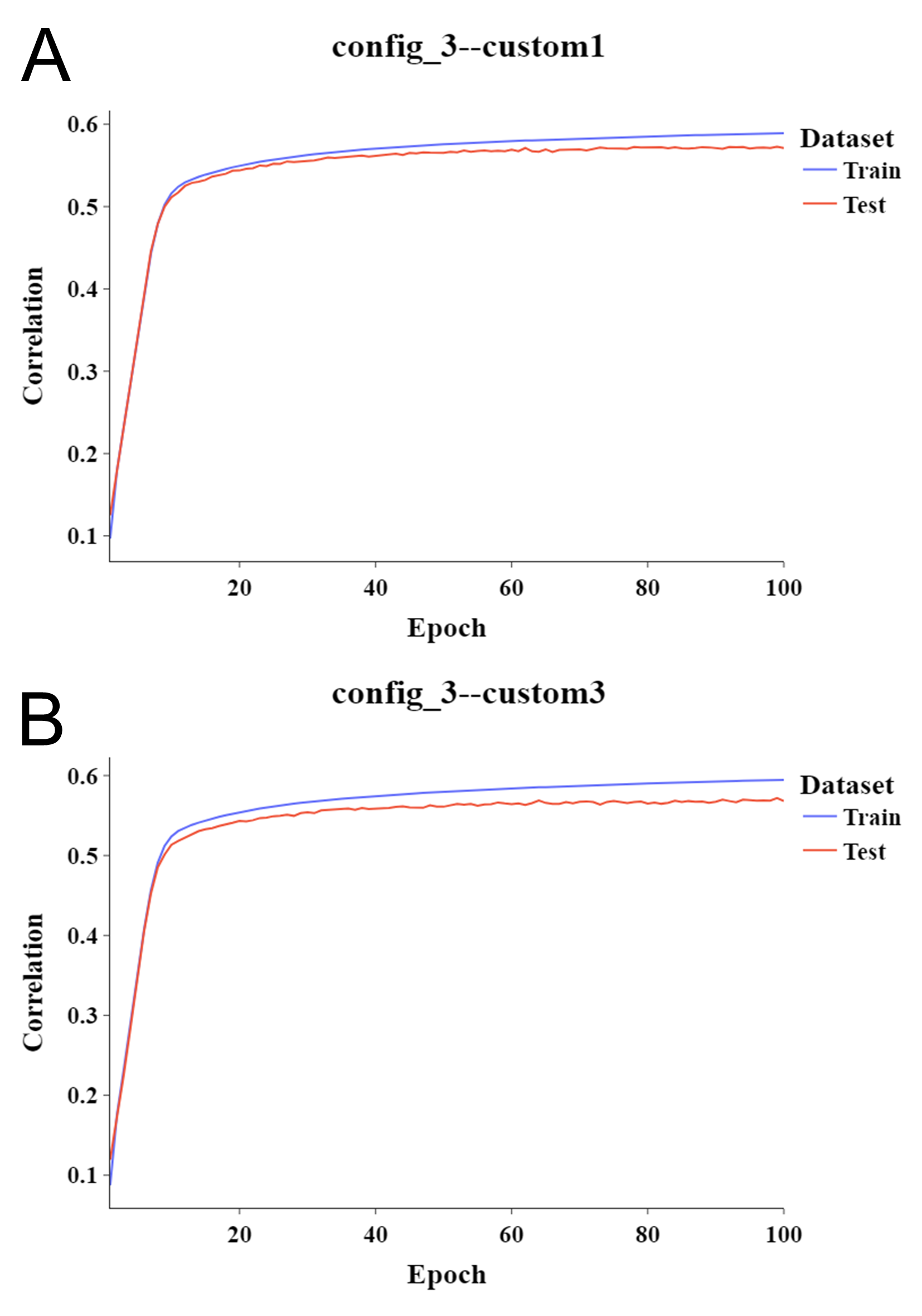}}
	\caption{{\bf Using encoder to build the final model by extending over a few CNN layers.}
		(A) Less complex model showing better learning curve. (B) A more complex model shows an inconsistent relation in training testing data.}
	\label{fig4}
\end{figure}

These results underscore the significance of incorporating nonlinear activation functions and batch normalization into our autoencoder architecture. AutoEncoder-3 emerged as the most promising choice, offering improved learning capabilities and generating high-quality images with enhanced details.

\subsection{Building the HistoSPACE model}

The HistoSPACE model uses the STNet Dataset, which has spatial expressions for corresponding H\&E images. These expression values will work as a supervision label for model training. To obtain our final model, we were required to add additional layers to the encoder to predict the spatial gene expression. Hence, we conducted a series of experiments to evaluate the impact of model complexity over performance, specifically by augmenting the number of layers in our architecture. We extended our baseline model with additional layers in our investigations, progressively increasing its complexity. To comprehensively assess these variations, we employed three distinct configurations, allowing us to gauge the influence of complexity on model performance under diverse circumstances.

The model was finalized based on its performance in terms of correlation. Our findings indicate that the less complex model consistently exhibits higher correlation and displays a more favorable convergence behavior (Figure \ref{fig4}), suggesting that an overly complex architecture may not necessarily yield improved results. These observations were further corroborated through testing across various configurations ({\bf See supplementary figure s2}) to obtain our final model with a single convolution block and two fully connected layers augmented in the image encoder. This final model, HistoSPACE, gives a correlation of 0.56 between predicted and actual expression values.

\subsection{Compare with the existing model}

We are comparing our model's performance with one of the foundational models, the STNet model,  for this task. We have performed a detailed comparison of model learning behavior and case-specific predicted results. STNet is a benchmark for assessing the progress and refinement of subsequent models developed in this domain. By comparing these two models, we gain valuable insights into the extent of performance enhancement achieved in our model.

\begin{figure}[H]
	\centerline{\includegraphics[width=\columnwidth]{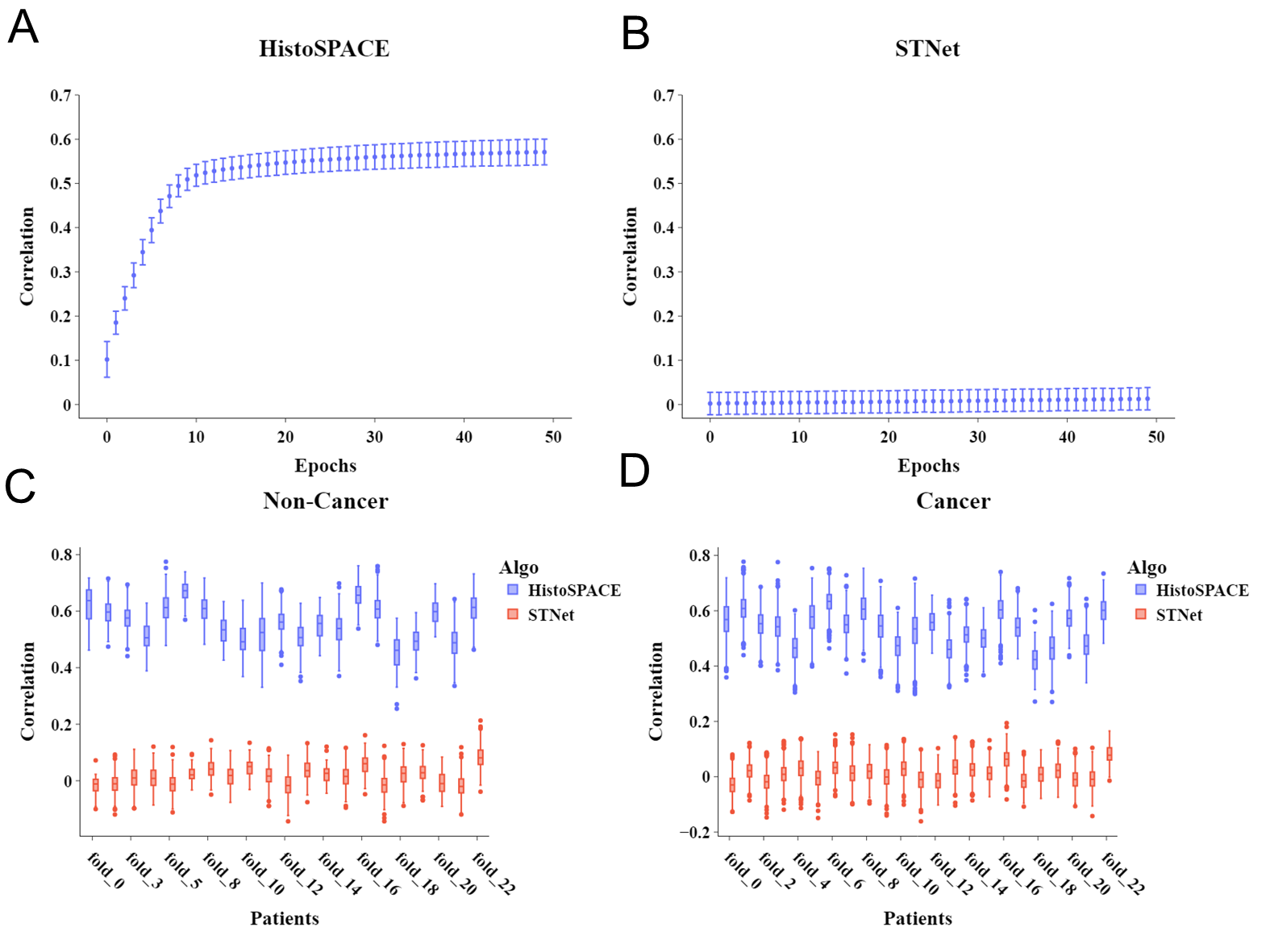}}
	\caption{{\bf Comparison of HistoSPACE with STNet with leaving one out validation scheme.}
		(A) Growing learning curve of STNet till 50 epochs. (B) Learning saturation achieved by our model in 50 epochs. (C-D) Expression distribution of both the algorithms for non-cancer/cancer tiles in particular samples.}
	\label{fig5}
\end{figure}

We observe the correlation between the predicted and actual values for the STNet model over 50 training epochs (Figure \ref{fig5}B). The gradual and linear increase in correlation indicates that the STNet model continues to learn from the dataset as training progresses.

The correlation trend for our model is shown in Figure \ref{fig5}A. Unlike STNet, correlating 0.01, our model exhibits a power-law-like correlation increase up to 0.56. It implies that the HistoSPACE model rapidly learns during the initial training epochs, reaching a high correlation relatively quickly. Subsequently, the model fine-tunes its performance at a slower rate. HistoSPACE has reached between 0.5 and 0.6, where additional training may yield diminishing or no improvements. This comparison with the STNet model suggests that our model has made substantial progress and possibly offers enhanced predictive capabilities for this dataset.

\begin{figure}[H]
	\centerline{\includegraphics[width=\columnwidth]{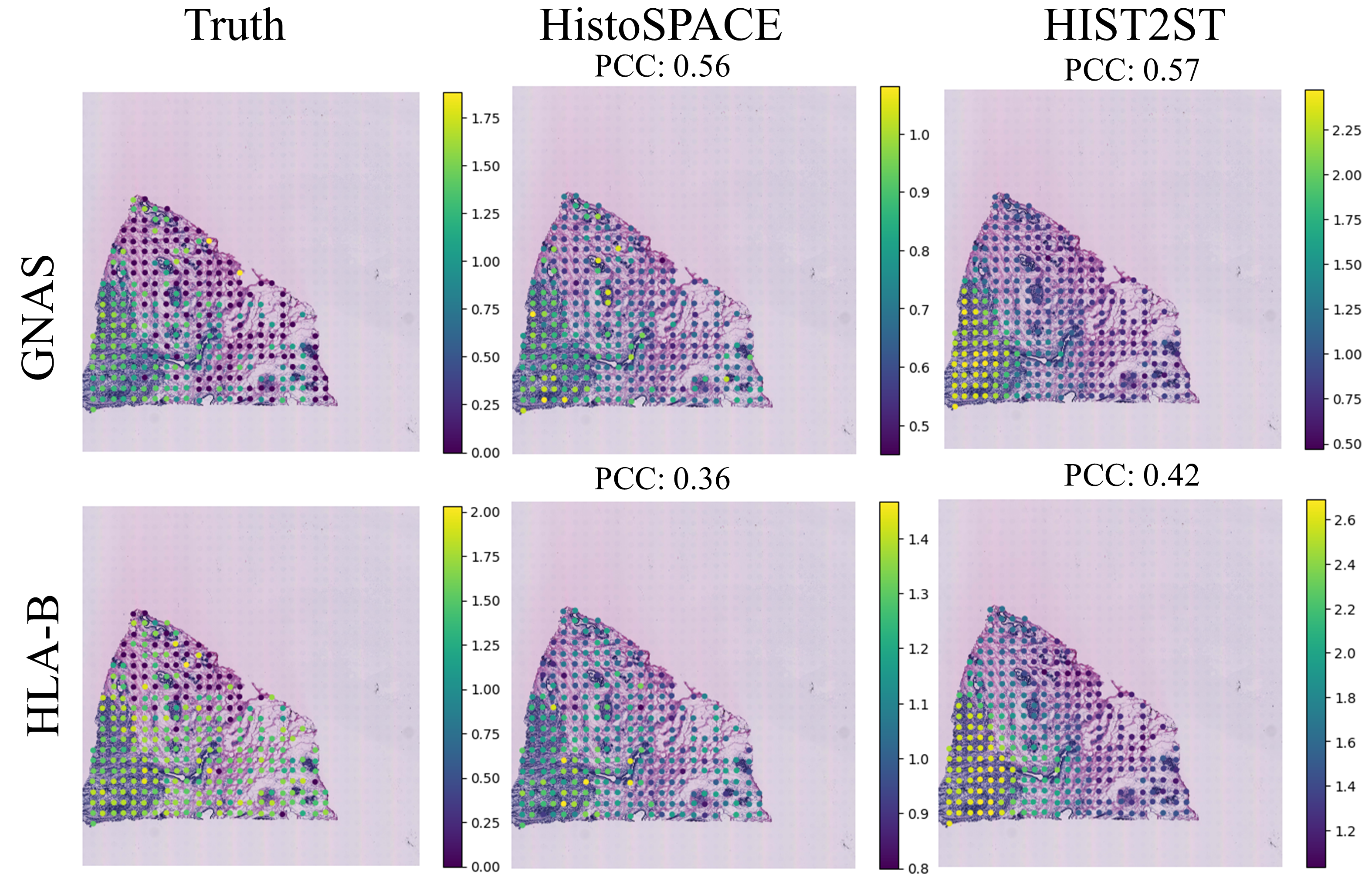}}
	\caption{Predicted expression on unseen data compared with other models trained on the same data. The truth column represents the actual expression for two gene GNAS and HLA-B.}
	\label{fig6}
\end{figure}

\subsection{Performance evaluation of HistoSPACE with novel data}

Spatial transcriptome profiles are unstable, which might lead to unexpectedly poor prediction results in unknown data. So, here, we are using the Her2 dataset to evaluate the robustness of our model. We have also compared our model with HIST2ST, built on this data, but our model has never seen this dataset. We are assessing the performance of our model in predicting gene expression patterns, particularly for genes GNAS and HLA-B, which are two high-performing genes. Our model achieved a correlation of 0.56 for GNAS and 0.36 for HLA-B compared to the reference model HIST2ST, which achieved correlations of 0.57 for GNAS and 0.42 for HLA-B on the same dataset.

\begin{figure}[H]
	\centerline{\includegraphics[width=0.6\columnwidth]{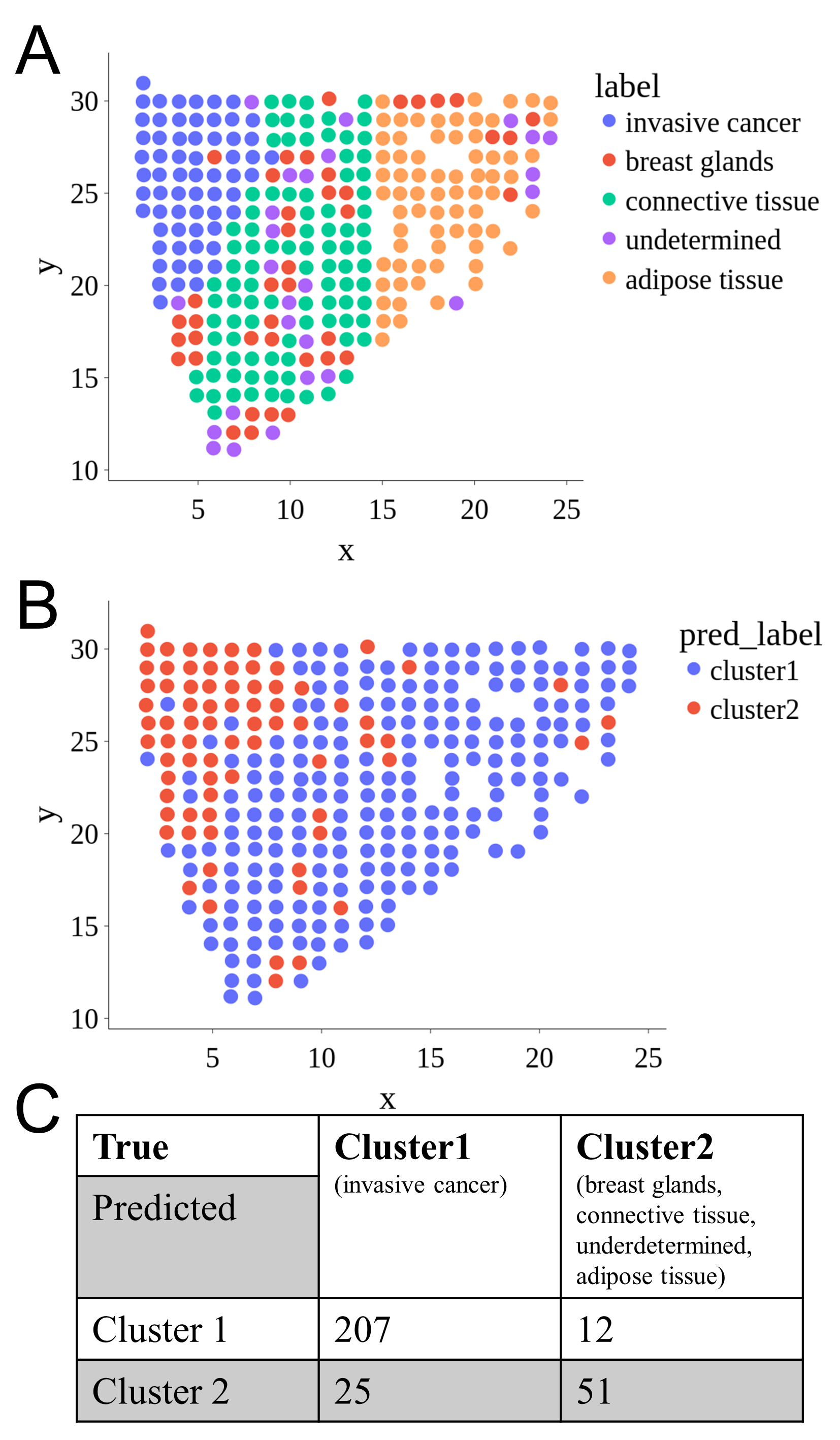}}
	\caption{{\bf The tissue physiology captured our algorithm compared to the ground truth.}
		(A) The true morphology annotated by a pathologist. (B) The distinction captured by predicted expression from our algorithm between cancer and non-cancer regions. (C) The truth table shows the number of elements are matched in original cluster and predicted clusters. The original annotation has been clubbed as cluster 1 for invasive cancer and cluster 2 for breast glands, connective tissue, undetermined and adipose tissue.}
	\label{fig7}
\end{figure}

Notably, both genes, GNAS and HLA-B, exhibited a strong positive correlation with our model, indicating the effectiveness of our approach in capturing and predicting gene expression profiles. Even more compelling is that our model demonstrated this high level of correlation without ever being exposed to the specific dataset used for HIST2ST training. To visualize the expression patterns, we plotted the results, and it is evident from the figures that our model's predictions closely align with those of the HIST2ST model, despite our algorithm being entirely unfamiliar with the dataset used for HIST2ST training. This observation underscores the ability of our model to generalize expression patterns to unseen and diverse datasets, showcasing its robustness and utility in predicting gene expression from histology images (Figure \ref{fig6}). We further investigated the capability of our algorithm to predict gene expression patterns while capturing the underlying tissue morphology, distinguishing between cancerous and non-cancerous regions. Our algorithm was good at predicting gene expression and demonstrated the capacity to construct the annotated tissue morphology. Our evaluation found that our model effectively discriminated between cancerous and non-cancerous tissue regions, aligning its predictions with the annotated morphology. This alignment was substantiated by our algorithm consistently associating specific gene expression patterns with the respective cancer and non-cancer regions (Figure \ref{fig7}). This observation underscores the potential of our model in capturing the spatial patterns exhibited by genes while accurately reflecting the underlying tissue context. By effectively discerning between cancerous and non-cancerous areas, our model holds promise as a valuable tool for dissecting intricate gene expression relationships within tissue morphology.

\section{Discussion}

In this study, we presented an algorithm designed for better expression prediction from histology images. Currently, spatial expression analysis is home to a limited number of existing algorithms, such as STNet \cite{he2020integrating} and Hist2ST \cite{zeng2022spatial}. However, these established algorithms exhibit predictive correlation as low as 0.1-0.3 and employ complex models such as pre-trained DenseNet121 \cite{huang2017densely} and transformer \cite{vaswani2017attention} emended with GNN. Assessing AI models on independent datasets is crucial to understanding how well they perform on other datasets \cite{geirhos2018generalisation}. In this context, it is essential to highlight that these models have yet to undergo testing on independent datasets, raising concerns about generalizing effectively. This compelling motivation led to the development of our model, HistoSPACE. Notably, HistoSPACE adopts a more streamlined model architecture, distinguishing itself by its simplicity, efficiency and robustness.

In practice, feature extraction is vital for image classification, but the image may only sometimes align directly with the output labels during supervised learning. This leads to a poorly trained model that needs to capture the right signal to make a good prediction on the test dataset. To enhance model performance, an independent feature extractor like autoencoder could be employed \cite{siradjuddin2019feature,meng2017relational}. This feature extractor effectively captures relevant features, as they are unsupervised and require no explicit supervised labels \cite{le2018supervised}. Leveraging the extracted features from an unsupervised task can be coupled with a supervised one for better performance with a more effective and accurate final model. This methodology will perform better because it does not need to extract explicit features from the images. 

HistoSPACE works on a similar principle as image regression is not a common supervision task. It begins with an image autoencoder designed to regenerate the image close to the original, with a deficient reconstruction error of 0.002. The autoencoder effectively regenerates images, remaining faithful to the original inputs. Subsequently, our core algorithm harnessed the encoder component as a feature extractor. The resultant model offers a correlation of 0.56 between predicted and actual expressions, surpassing the performance of STNet, the precursor algorithm associated with this dataset.

We extended our testing on an independent dataset, which is used to develop Hist2ST \cite{zeng2022spatial}. The performance of our model on this new data remained the same compared to our training data, demonstrating its robustness. It should be noted that its performance was at par with Hist2ST despite not being trained on that specific dataset. Our final trained model is 30 times smaller in storage size than Hist2ST. HistoSPACE is a comparatively simpler model with good and robust performance, promising its potential to be applied to different datasets as a general model.

As a final step, we explored its possible application toward mechanistic understanding driven by image-based tasks. We formed two clusters using predicted expression values from our HistoSPACE model. Then we compared this cluster with true annotations of cancers and non cancer regions performed by the pathologists. We found that out of 295 elements available in both the clusters, we were able to capture 258 element successfully. Notably, our algorithm demystifies images in terms of true expression, which offers a mechanistic approach to classifying cancer and non-cancer tiles, providing a distinct perspective from conventional image classification models. This thorough evaluation established the robustness and versatility of HistoSPACE in handling independent datasets and its potential for diverse biomedical applications.

Before ending the article, we would like to mention that HistoSPACE outperforms peer models in terms of simplicity and generalization, but it has yet to undergo evaluation for organ-specific or cross-organ expression prediction. This avenue holds promise for future exploration. We anticipate that, notwithstanding its current limitations, our algorithm will pave the way for innovative advancements in precision medicine.

\section*{Acknowledgment}

Research of SK is supported by Translational Health Science and Technology Institute (THSTI) Ph.D Fellowship.

\bibliographystyle{}
\bibliography{ref}

\end{document}